\documentclass[preprintnumbers,amsmath,amssymb,prd,superscriptaddress,notitlepage]{revtex4-1}

\usepackage{graphicx}
\usepackage{dcolumn}
\usepackage{bm}

\newcommand{\be}{\begin{equation}}
\newcommand{\ee}{\end{equation}}

\newcommand{\vx}{\mathbf{x}}

\newcommand{\vy}{\mathbf{y}}

\newcommand{\mnras}{{\em Mon. Not. Roy. Astron. Soc. }}

\newcommand{\jcap}{JCAP}
\newcommand{\physrep}{{\em Phys. Rept. }}
\newcommand{\aap}{{\em Astron. Astrophys. }}

\newcommand{\pasj}{PASJ}

\def\prd{{\em Phys. Rev. }{\bf D }}

\newcommand{\dd}{\delta^\prime}

\begin{document}

\title{What the ``simple renormalization group'' approach to dark matter
clustering really was}

\author{Patrick McDonald}
\email{PVMcDonald@lbl.gov}
\affiliation{Lawrence Berkeley National Laboratory, One Cyclotron Road,
Berkeley, CA 94720, USA}
\affiliation{(formerly) Brookhaven National Laboratory, Building 510A,
Upton, NY 11973-5000, USA}

\date{\today}

\begin{abstract}

McDonald (2007) presented an approach to improving perturbation theory (PT)
calculations of the dark matter power spectrum, with a derivation based on the 
idea of renormalization group flow with time. In spite of a questionable 
approximation made in deriving it, subsequent comparisons by several groups 
between the predictions of the resulting equation and N-body simulations showed
remarkable improvement relative to ``standard'' PT (SPT) at similar order. In 
this brief note I show that the same final equation can be derived cleanly from
the point of 
view not of flowing with time but with non-linear coupling strength, i.e., 
gradually dialing the coupling from the trivial value zero to the physical one.
This understanding makes it clear how to extend the approach to higher order 
and other statistics. While I do not necessarily think this approach
is best among the many, it may be interesting in that it contains 
a unique way of suppressing UV sensitivity.  In passing 
I remind the reader of references demonstrating that SPT works remarkably well 
without improvement (except near redshift zero, where, fortunately for
SPT, there is very little volume in the Universe). 

\end{abstract}

\maketitle

\section{Introduction}

Recently there has been a lot of work toward improving perturbative predictions
for dark matter clustering \cite{2007PhRvD..75d3514M,
2012ApJ...760..114S,2012PhRvD..86j3528T,
2012JHEP...09..082C,2014PhRvD..89d3521H,2012JCAP...06..021R,
2012JCAP...06..018R,2012JCAP...03..031S,2012JCAP...01..019P,
2011JCAP...06..015A,2011JCAP...04..032M,
2009PhRvD..80l3503T,2009PhRvD..80d3504P,
2009PASJ...61..321N,2008PhRvD..78j3521B,
2008PhRvD..78h3503B,
2008JCAP...10..031N,
2008A&A...484...79V,2008PhRvD..77f3530M,2008ApJ...674..617T,
2008PhRvD..77b3533C,2007MNRAS.382.1460P,
2007JCAP...06..026M,2007A&A...465..725V}. 
One could argue that 
this is an academic indulgence, because we can use N-body simulations to 
calculate
the same results essentially exactly from first principles; however, we will 
never be able 
to simulate galaxies and other tracers of dark matter anywhere near exactly 
from first 
principles, so in my opinion it is vital to develop a deeper 
understanding of what forms of large-scale clustering are generally possible
\cite{2009JCAP...08..020M,2006PhRvD..74j3512M,2013arXiv1310.2672N,
2012PhRvD..86j3519C,
2011MNRAS.417L..79G,
2011PhRvD..83h3518M,2010MNRAS.408.2397M,2010PhRvD..82d3515H,
2010MNRAS.406..803B,2010PhRvD..81j3527M,
2009JCAP...11..026M,2010PhRvD..81f3530G,
2009PhRvD..80j5005E,2009PhRvD..80h3528S,2009ApJ...693.1404S,
2008PhRvD..78l3519M,2009ApJ...691..569J,2008PhRvL.100i1303C,
2007PhRvD..76h3005L,2007PhRvD..76f3009M,2003ApJ...585...34M}
The power of future high precision large-scale structure surveys to constrain 
fundamental physics will depend
sensitively on how well we can model these tracers 
\cite{2013arXiv1308.4164F,2009JCAP...10..007M}.
Understanding
dark matter clustering beyond what is possible by running simulations is a
prerequisite for more general understanding of large-scale structure. 

In \cite{2007PhRvD..75d3514M} I presented one relatively early approach to 
improving ``standard''
(in the sense of being what people have done for a long time 
\cite{1980lssu.book.....P,1981MNRAS.197..931J,1983MNRAS.203..345V,
1984ApJ...279..499F,1986ApJ...311....6G,2002PhR...367....1B})
Eulerian perturbation theory (SPT). 
The calculation in \cite{2007PhRvD..75d3514M} was motivated by the general
renormalization group approach of
\cite{1994PhRvL..73.1311C}, and explained as an implementation of that method;
however, a complete implementation of the method of \cite{1994PhRvL..73.1311C}
would have required non-negligible work beyond standard perturbation theory.
Finding that a
corner-cutting version of the calculation seemed to work well by comparison to
simulation fitting formulas, \cite{2007PhRvD..75d3514M} swept the conceptual
deficiencies under the rug and declared victory. Specifically:
the method of \cite{1994PhRvL..73.1311C} is essentially to reset the initial 
conditions (IC) for perturbation theory at a 
continuous series of time steps, absorbing the corrections in the previous step 
into the initial conditions for the next step. Implementing the method properly
would have
taken more work because the corrections don't take the form of
growing modes, but standard PT is written for growing mode initial conditions.
To be clearer: at 2nd order in PT $\delta$ and $\theta$ are not simply equal
to each other as they would be for linear theory growing modes (when 
appropriately defined).
This means that the 2nd order results cannot be simply plugged back in place
of the usual linear theory initial conditions -- to do this properly requires
deriving the PT evolution for a short time after an arbitrary start (normally
the use of only the fastest growing modes is justified by the fact that the
starting point is effectively infinitely far in the past).  This is not 
actually very hard, but \cite{2007PhRvD..75d3514M} took the easy way out after
the embarrassingly poorly motivated ``approximation'' of ignoring the
sub-leading modes appeared to succeed (if it is not obvious why we shouldn't
expect this to be a 
reasonable approximation, it is because we know that a non-negligible part of
the SPT prediction flows for some time through these modes -- it is really
only in the infinite-past initial conditions that they are negligible). 

\cite{2007PhRvD..75d3514M} was aware of the deficiencies of the calculation and
suggested that one way forward might be to simply write down and numerically 
evolve the full evolution 
equations for statistics, i.e., the power spectrum, bispectrum, trispectrum, 
etc., an idea that was implemented as the ``time-RG'' method by 
\cite{2008JCAP...10..036P}. 
However, predictions
of the equations of \cite{2007PhRvD..75d3514M} were subsequently compared in 
more detail to results from N-body simulations by several groups, finding 
remarkable improvement over standard perturbation theory, for a diverse set
of scenarios. 
\cite{2009MNRAS.397.1275W} simulated a wide range of power law initial 
conditions, from $n=-1$ to $n=-2.5$, finding (their Fig. 5) that the RG 
predictions
were always a substantial improvement over the 1-loop SPT result.
\cite{2009PhRvD..80d3531C} did not include the method of 
\cite{2007PhRvD..75d3514M}, which they call RGPT, in most of their comparisons
(possibly because I was telling them at the time that they should consider 
``time-RG'' to be the natural evolution); however, where they
do show it (the cyan line in their Fig. 3) we see excellent agreement at 
$z=1$.
Basically, we see improvements comparable or somewhat better than going from 
1-loop to 2-loop SPT (which, as we see in Fig. 1 of \cite{2009PhRvD..80d3531C}
is impressive in itself
at $z=1$ -- I will discuss this, including redshift dependence, more below), 
so we can either be happy not to 
need to deal with the pain of the
2-loop calculation, or we can hope for better by taking the RG method to
higher order. 
Finally, \cite{2011PhRvD..84f3501O} present simulations of an interesting 
combination of different power laws (from $n=-0.5$ to $-1.5$) with a
BAO-like feature, showing (their Fig. 11) that ``coupling strength'' RGPT 
(named based on a draft of this note) performs well. 
To be fair, the results in these papers are far from perfect, but they show 
that the 
method generally does what it is supposed to, improve SPT, and does it 
arguably as well as any others \cite{2009PhRvD..80d3531C}. It may be that other
methods derived more recently are better, but maybe we just needed to work 
harder on this one. 
This motivates this note in which I show how to derive
the results of \cite{2007PhRvD..75d3514M} without hacking, and then how to 
extend it to higher order and other statistics. 

\section{Re-derivation and interpretation of the ``Simple RG" result }

The basic form of the SPT solution, defining $A=D^2$, is:
\begin{equation}
P(k,A,\lambda) = A P_L(k) + \lambda A^2 [P_L^2](k) +
\lambda^2 A^3[P_L^3](k)+...
\end{equation}
where $[P_L^n]$ are shorthand for the convolution of $n$
power spectra that one finds in the $n$th order SPT calculation, and $\lambda$
is a non-linear coupling strength parameter.
I have written $P(k,A,\lambda)$ to emphasize that I have generalized the 
solution to apply for any coupling strength, including $\lambda=0$, where linear
theory works perfectly by definition, and the true physical value $\lambda=1$. 
The strategy here will be to use an RG approach to ``turn on'' the non-linear 
coupling, in some sense evolving the solution from the known starting point 
$\lambda=0$ to the point of interest $\lambda=1$. Because of the simple form
taken by time evolution in LSS PT, this approach looks very much like a time
evolution, but that is really not what we are doing here. 

The key to renormalization is to exploit a general ambiguity inherent in 
perturbation theory, that the perturbative ``order'' of different quantities in
the calculation
is not god-given --  it is up to the user to define things in a way that gives
the best possible behavior of the resulting series. 
The ``standard'' LSS perturbation theory calculation assumes that the initial 
conditions $P_L(k)$ go
entirely into the lowest order solution, but this is not mathematically
necessary. If we allow for IC at each order, $P_1(k)$, $P_2(k)$, and $P_3(k)$,
we find 
\begin{equation}
P(k,A,\lambda) = A \left(P_1\left(k\right)+P_2\left(k\right)+
P_3\left(k\right)\right) 
+\lambda A^2\left(\left[P_1^2\right]\left(k\right)+2\left[P_1 P_2\right]
\left(k\right)\right)
+\lambda^2 A^3\left[P_1^3\right]+...
\end{equation}
where note that the exact same solutions apply regardless of the order label on
the initial conditions -- the label just determines which terms to drop
(to be clear, $[P_1 P_2]$ is the usual 1-loop convolution using $P_1$ for
one of the power spectra and $P_2$ for the other, symmetrically of course). 

For any given value of $\lambda$, call it $\lambda_\star$, we can specify
$P_2$ and $P_3$ to make the beyond-linear corrections zero, i.e., 
\begin{equation}
P_2(k,\lambda_\star)=-\lambda_\star \left[P^2_1(\lambda_\star)\right]
\end{equation}
where the leading order result has become a function of $\lambda_\star$, 
$P_1(k,\lambda_\star)$, and
\begin{equation}
P_3(k,\lambda_\star)=
-2 \lambda_\star \left[P_1(\lambda_\star)P_2(\lambda_\star)\right]
-\lambda_\star^2 \left[P^3_1(\lambda_\star)\right]~.
\end{equation}
We suppress the $A$ dependence of everything in these equations as it plays
no role (one should think of the calculation as being done at some fixed $A$). 
The complete solution as a function of $\lambda$ is then
\begin{equation}
P(k,\lambda) = P_1\left(k,\lambda_\star\right)+
(\lambda-\lambda_\star) 
\left(\left[P_1^2(\lambda_\star)\right]\left(k\right)+
2\left[P_1(\lambda_\star) P_2(\lambda_\star)\right]
\left(k\right)\right)
+(\lambda^2-\lambda_\star^2) \left[P_1^3(\lambda_\star)\right]+...
\label{eq:Poflambdastar}
\end{equation}
Now, as usual in RG calculations, we enforce the fact that the solution should
not depend on the arbitrary cancellation point $\lambda_\star$, i.e., the 
derivative of the solution, $P(k,\lambda)$, as laid out in equation 
\ref{eq:Poflambdastar},
with respect to $\lambda_\star$ should be zero. This should be true at any
value of $\lambda_\star$ but
for convenience we take this derivative at $\lambda_\star=\lambda$, i.e., 
set 
$\left. \frac{dP(k,\lambda)}
{d\lambda_\star}\right|_{\lambda_\star=\lambda}= 0$,  to derive:
\begin{equation}
\frac{dP(k,\lambda)}{d\lambda}=
\left[P^2(\lambda)\right]\left(k\right)
+2 \lambda \left(\left[P^3(\lambda)\right]-
\left[P(\lambda)\left[P^2(\lambda)\right]\right]\right)+...
\label{eq:basicRG}
\end{equation}
where I have used the fact that $P(k,\lambda)=P_1(k,\lambda)$ and
$P_2(k,\lambda)=-\lambda \left[P_1^2(\lambda)\right]$, when 
$\lambda_\star=\lambda$. 

The first term here is the result of \cite{2007PhRvD..75d3514M}. The 2nd
shows how to go to the next order. To be clear, 
$\left[P^2(\lambda)\right]\left(k\right)$ is the usual 1-loop SPT calculation,
with $P(k,\lambda)$ as the input power spectrum,
$\left[P^3(\lambda)\right](k)$ is the usual 2-loop SPT calculation, and
$\left[P(\lambda)\left[P^2(\lambda)\right]\right](k)$ means do the 1-loop SPT 
convolution with one of the two input power spectra (symmetrically) 
replaced by the 1-loop 
result itself. The initial condition for solving the differential equation is
$P(k,\lambda=0)=P_L(k)$, i.e., at zero non-linear coupling the solution is
obviously the linear solution.  

\section{Extension to other quantities}

It should be clear at this point how to extend this calculation to other
statistics. For example, redshift space distortions in the distribution 
function approach 
\cite{2013arXiv1312.4214O,2013JCAP...10..053V,2012JCAP...11..014O,
2012JCAP...11..009V,2012JCAP...02..010O} require predictions for statistics
involving momentum, motivating computing $\dot{\delta}\equiv d\delta/dt$
which is equivalent to the divergence of momentum  
through the continuity equation. I also show how to compute the propagator 
of \cite{2006PhRvD..73f3520C,2006PhRvD..73f3519C}. 

\subsection{$P_{\delta\dot{\delta}}$}

We can compute $P_{\delta\dot{\delta}}=\frac{1}{2}
dP_{\delta\delta}/dt$ by evaluating Eq. \ref{eq:basicRG} at two nearby times
to take a numerical derivative, or directly through:
\begin{equation}
\frac{d P(k,A,\lambda)}{d\ln A}\equiv P^\prime(k,A,\lambda) = 
A \left(P_1\left(k\right)+P_2\left(k\right)+
P_3\left(k\right)\right) 
+2 \lambda A^2\left(\left[P_1^2\right]\left(k\right)+2\left[P_1 P_2\right]
\left(k\right)\right)
+3 \lambda^2 A^3\left[P_1^3\right]+...
\end{equation}
and then, following the same calculation as above,
\begin{equation}
P_2(k,\lambda_\star)=-2 \lambda_\star \left[P^2_1(\lambda_\star)\right]
\end{equation}
\begin{equation}
P_3(k,\lambda_\star)=
-4 \lambda_\star \left[P_1(\lambda_\star)P_2(\lambda_\star)\right]
-3 \lambda_\star^2 \left[P^3_1(\lambda_\star)\right]
\end{equation}
\begin{equation}
P^\prime(k,\lambda) =  P_1\left(k,\lambda_\star\right)+
2 (\lambda-\lambda_\star) 
\left(\left[P_1^2(\lambda_\star)\right]\left(k\right)+
2\left[P_1(\lambda_\star) P_2(\lambda_\star)\right]
\left(k\right)\right)
+3 (\lambda^2-\lambda_\star^2) \left[P_1^3(\lambda_\star)\right]+...
\end{equation}
\begin{equation}
\frac{dP^\prime(k,\lambda)}{d\lambda}=
2 \left[P^{\prime 2}(\lambda)\right]\left(k\right)
+2 \lambda \left(3 \left[P^{\prime 3}(\lambda)\right]-
4 \left[P^\prime(\lambda)\left[P^{\prime 2}(\lambda)\right]\right]\right)+...
\end{equation}
using $P^\prime(k,\lambda)=P_1(k,\lambda)$. While the form is similar to 
equation \ref{eq:basicRG}, there is really no simple analytic relation between
$P^\prime(k,A,\lambda)$ and $P(k,A,\lambda)$. 

\subsection{$P_{\dot{\delta}\dot{\delta}}$}

If we are interested in computing $P_{\dot{\delta}\dot{\delta}}$, we must go 
back a step, as this cannot be written as a simple derivative of the power
spectrum. The PT density field is 
\begin{equation}
\delta(k,D,\epsilon)=D \delta_L(k)+D^2\epsilon [\delta_L^2](k)
+D^3\epsilon^2 [\delta_L^3](k)+...
\end{equation}
\begin{equation}
\frac{d\delta(k,D,\epsilon)}{d\ln D}=
D \delta_L(k)+2 D^2\epsilon [\delta_L^2](k)+ 3 D^3\epsilon^2 [\delta_L^3](k)+...
\end{equation}
or (tracking only 1 loop here for simplicity)
\begin{equation}
P_{\delta^\prime \delta^\prime}=A P_L(k)+
4 A^2 \lambda \left[P_L^2\right]_{22}(k)+
6 A^2 \lambda \left[P_L^2\right]_{13}(k)+...
\end{equation}
where we have defined $\left[P^2\right]_{22}$ and
$\left[P^2\right]_{13}$ such that the standard PT density power spectrum
result would be $P_L+\left[P_L^2\right]_{22}+2\left[P_L^2\right]_{13}$ 
($P_{13}$ may sometimes be defined to include the factor of 2).
Defining $\left[P^2\right]_{\delta^\prime \delta^\prime}=
4 \left[P^2\right]_{22}(k)+
6 \left[P^2\right]_{13}(k)$ 
we have
\begin{equation}
P_{\delta^\prime \delta^\prime}=A(P_1(k)+P_2(k))
+\lambda A^2 \left[P_1^2\right]_{\delta^\prime \delta^\prime}
\end{equation}
i.e., 
\begin{equation}
P_2(k,\lambda_\star)=-
\lambda_\star \left[P_1^2(\lambda_\star)\right]_{\dd\dd}
\end{equation}
i.e., 
\begin{equation}
\frac{dP_{\dd\dd}(k,\lambda)}{d\lambda}=
\left[P_{\dd\dd}^2(\lambda)\right]_{\dd\dd}\left(k\right) ~.
\label{eqprimeprimeRG}
\end{equation}
The equation to be solved again has similar form but non-trivially different
coefficients. 

\subsection{Propagator}

Another thing we can compute is the propagator, 
$P_{\delta \delta_L}/P_L$, i.e., the standard PT result is
\begin{equation}
P_{\delta \delta_L}(k,A,\lambda)=A P_L(k)+A^2 \lambda\left[P_L^2\right]_{13}
\end{equation}
and we see immediately that the RG equation is 
\begin{equation}
\frac{dP_{\delta\delta_L}(k,\lambda)}{d\lambda}=
\left[P_{\delta\delta_L}^2(\lambda)\right]_{13}\left(k\right)
\label{eq:propagatorRG}
\end{equation}
where again note that $\left[P^2\right]_{13}$ is defined to account for only
a single product $\left<\delta_1\delta_3\right>$. 

\section{Discussion}

It should be clear from these examples how to derive any other desired 
quantity. E.g., 
$P_{\theta\theta}$ would follow equation (\ref{eq:basicRG}) with the relevant 
kernels for $\theta$ replacing those for $\delta$. 
I am probably setting a bad example with these extensions, however, as it is 
probably best to use this RG method only to compute complete observables
of interest, not to compute sub-components of observables separately with the 
intention of adding them, because 
that hides potentially exact cancellations from the method. 
This is exemplified by comparing the result for the propagator,
equation (\ref{eq:propagatorRG}) to the full power spectrum, equation 
(\ref{eq:basicRG}). We can see looking at equation (\ref{eq:propagatorRG}) 
that it will
produce the expected behavior of a propagator, as discussed in detail by 
\cite{2006PhRvD..73f3520C,2006PhRvD..73f3519C}. Because of the form of
$[P^2]_{13}(k)$, proportional to $P(k)$, we could re-write equation 
(\ref{eq:propagatorRG}) as an equation for 
$\frac{d\ln P_{\delta\delta_L}(k,\lambda)}{d\lambda}$ so that,
considering the general negativity of $[P^2]_{13}(k)$, it is clear that 
$P_{\delta\delta_L}$ will stay positive but be driven to zero at high $k$, 
as we expect for the propagator. \cite{2006PhRvD..73f3520C,2006PhRvD..73f3519C}
presented this propagator as a fundamental building block of the calculation of
the full power spectrum, but this turns out to be problematic in principle
(although maybe not always in practice).
The problem is that the propagator is sensitive to 
large-scale bulk flows, with one well-known symptom being that $[P^2]_{13}$ is 
infrared divergent for sufficiently red power spectra, when the
velocity variance is also divergent. This is a physically correct 
result as far as the propagator goes -- a large homogeneous
velocity field really will decouple the initial conditions from the final 
field in fixed Eulerian coordinates, as defined by $P_{\delta\delta_L}(k)$, 
and the RG equation (\ref{eq:propagatorRG}) will do the right thing by sending 
the propagator to zero instead of negative infinity (for any reasonable
regulation of the divergence). 
However, it is easy to see that this potential infrared divergence is not 
relevant for the autocorrelation of the evolved field, or any correlation of
fields that are moving together with the bulk flow, because of Galilean 
invariance. Mathematically, this irrelevance of the potential infrared
divergence in
$P_{13}$ is correctly manifested in the SPT calculation of the autocorrelation
because the $P_{13}$ divergence exactly cancels against a similar 
divergence in $P_{22}$. 
The RG equation (\ref{eq:basicRG}) for the autocorrelation ``understands'' this
exact cancellation and therefore behaves fundamentally differently from the
combination of separate $P_{13}$ and $P_{22}$ calculations. While
$P_{13}$ and $P_{22}$ do not actually diverge for the 
$\Lambda$CDM power spectrum, the same physical intuition should apply more 
generally: it is clearly desirable to apply renormalization to fully 
observable quantities, where symmetry-enforced exact cancellations are 
manifest.

\begin{figure}[tb]
\centering
\includegraphics[width=\textwidth]{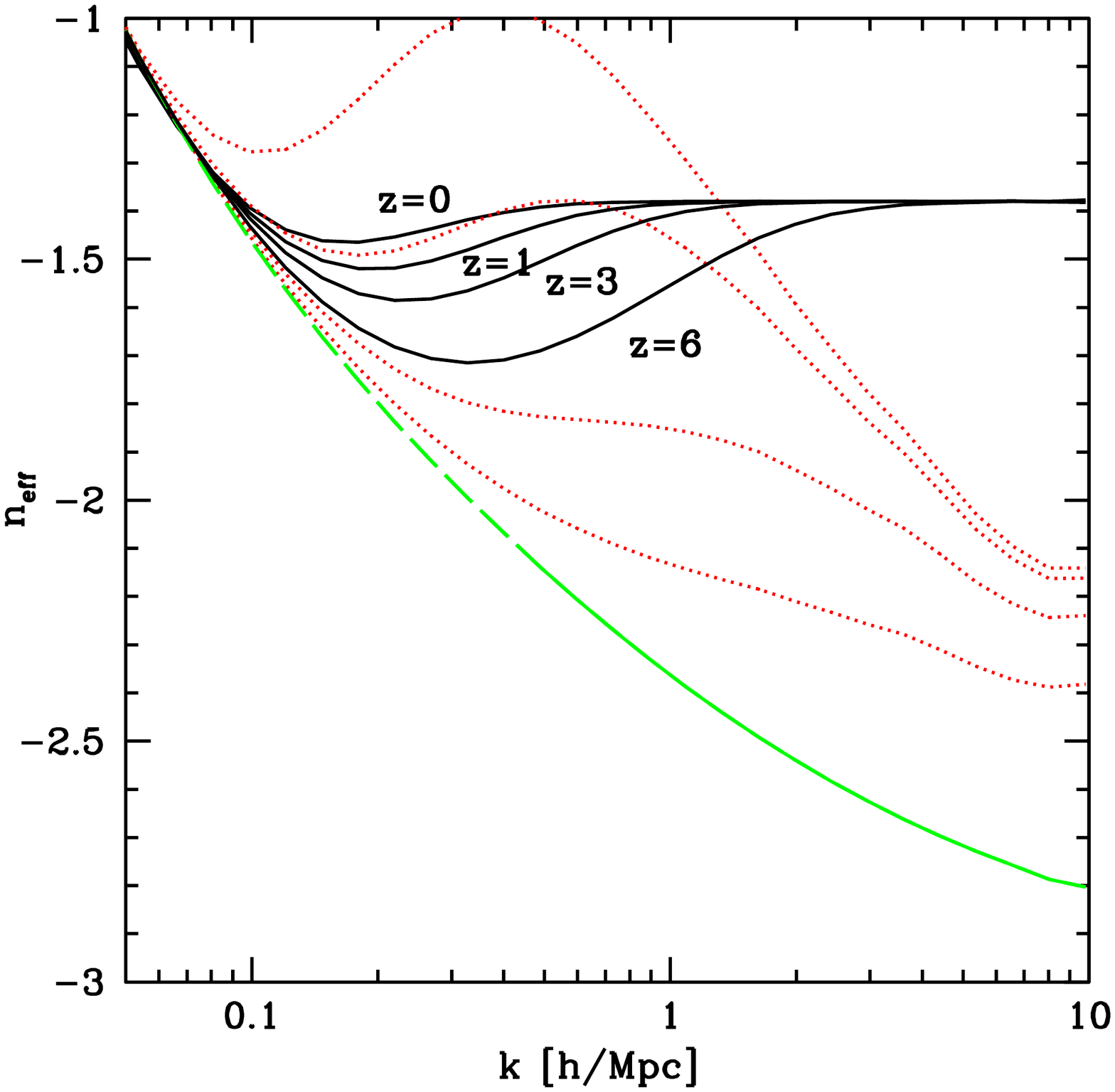}
\caption{
The green (dashed) line shows 
$n_{\rm eff}(k)\equiv \frac{d\ln P}{d\ln k}(k)$ for the linear power spectrum.
Black (solid) lines show $n_{\rm eff}(k)$ for the RG power spectrum 
(equation \ref{eq:basicRG}, at lowest order) at 
redshifts from top to bottom $z=0, 1, 3, 6$.
Red (dotted) lines show $n_{\rm eff}(k)$ for the 1-loop SPT power spectrum at 
the same redshifts. 
}
\label{fig:neff}
\end{figure}

As discussed in \cite{2007PhRvD..75d3514M}, the RG equation (\ref{eq:basicRG})
has another interesting property related to the fact that it combines 13 and
22 terms -- it has a fixed-point solution (at 1-loop order) at a
$k^{-1.4}$ power spectrum, i.e., the power spectrum where 1-loop SPT 
corrections have been shown to be exactly zero \cite{1996ApJ...473..620S}.
This is an attractive fixed point, i.e., a general power spectrum evolving 
under equation (\ref{eq:basicRG}) converges toward it, most quickly on small
scales, as shown in Figure \ref{fig:neff}.
This is interesting in that it erases sensitivity to small-scale details 
in a really unique way -- producing universal non-trivial small-scale 
behavior that is grafted onto the model-dependent large-scale
perturbative behavior. This differs qualitatively from something like the
propagator renormalization approach, which completely erases small-scale 
structure at lowest order and then needs to build it back up through 
loops \cite{2006PhRvD..73f3520C,2006PhRvD..73f3519C}, or the effective
field theory (EFT) approach 
\cite{2012JHEP...09..082C}
which removes small-scale sensitivity at the expense
of introducing a free parameter representing an effective sound-speed or 
viscosity for the effective fluid. While I generally like the EFT approach,
this raises the question of whether or not
that kind of free parameter is really necessary. It is easy to see situations
in which it unquestionably would be: 
if the system under discussion contained fundamentally
small-scale, highly non-trivial physics, e.g., star formation and supernova
explosions setting the temperature of some gas, obviously one is going to need
to represent that temperature by a free parameter. This is pretty close to 
exactly the situation with tracers of dark matter like galaxies, so an EFT
approach with free parameters is pretty obviously the best way to think about 
biasing 
\cite{2009JCAP...08..020M,2014arXiv1402.5916A}. 
On the other hand, is it possible that in certain simpler situations like the 
clustering of dark matter, where there are important conservation laws at work,
that the EFT parameters are more a property of the ``boundary layer''
non-linearities, accessible to perturbative calculations near that boundary,
instead of fundamentally sensitive to the initial conditions and things 
well below the non-linear scale? 
I have in mind a picture in which the 
formation of halos is like the closing of a box around the interior
of the halo, with the large-scale clustering subsequently only dependent on 
simple properties of these boxes, not what is happening inside them,
and with those properties determined by the structure on the presently 
collapsing 
scale, not what happened earlier inside. A simple example is to 
imagine a pair of sub-objects tightly orbiting each other inside the halo -- 
clearly you can make the binary tighter and change the kinetic energy in the
halo, but this isn't going to have any effect on the large-scale clustering of 
the halo, and therefore you don't need to be able to predict it (on the other
hand, the details of this binary definitely would affect the redshift-space 
power spectrum, so this thought experiment implies the need for a free 
parameter there, even for something as simple as dark matter).
This picture
of universal halos clustering independent of their internal details obviously
isn't anything new
\cite{2000MNRAS.318..203S,2000ApJ...543..503M,1997MNRAS.284..189M}, 
the question is just, does it imply that EFT parameters
could be predictable from first principles? The attractive fixed-point 
behavior seen in Figure \ref{fig:neff} suggests a way that this kind of
predictable self-arrangement of the EFT could work...  

To randomly change the subject, I'd like to comment on the relation of the 
power spectrum and its prediction to the correlation function. 
The most proximate motivation for this is the horrible jagged line
shown in Fig. 14 of \cite{2011PhRvD..84f3501O} as the transformation of the 
RGPT power spectrum prediction to a correlation function (accompanied by the
statement that they couldn't reasonably transform the SPT power spectrum at 
all), but I think there
is more generally a lot of confusion about this issue. 
The first thing one usually thinks is ``$P(k)$
and $\xi(r)$ are just linear transforms of each other, so they really 
should be equivalent'' which is then quickly followed by ``but this is
only true if one considers all values of $k$ and $r$, which is generally 
impractical, so really they can be different.'' I think one should look very
skeptically at any conclusion that they are different in an important way.
The first thing to note is that the naive correlation function 
$\left<\delta(\vx) \delta(\vy)\right>$ is generally a math/physically 
pathological statistic, because these $\delta$'s are defined at mathematical
points, and the statistic is therefore generally sensitive to arbitrarily 
small-scale physics. This is easy to see by considering the correlation 
function of a regular lattice of point-like objects, which will have delta
function spikes at multiples of the lattice spacing. Clearly the form of
these spikes is sensitive to the detailed small-scale structure of the objects,
no matter how large the separation in the correlation function. If we do  
not want the correlation function to be sensitive to small scale structure 
in this way, we need to apply some smoothing, enough to smear out the details
of the point objects. This smoothing is obviously equivalent to restricting
ourselves to considering the low-$k$ part of the power spectrum. This 
general small-scale sensitivity of the un-smoothed correlation function is 
obscured in 
cosmology because the small-scale structure is not arranged on a regular
grid, so it is naturally smeared, i.e., there is no significant small-scale
structure in the correlation function at large separations. The BAO feature
is a notable exception to this, where the discussion above explains why it is
affected by non-linearities that one would not notice at similar separation
if the feature was not present. The relevance to perturbation theory 
predictions is: it is a mistake to ask any method to predict the un-smoothed
correlation function. The correct procedure is to ask: ``what level of smoothing
can I apply to the correlation function without erasing cosmological structure 
that I believe to be present and predictably relevant to my cosmological 
parameter estimation,'' and then consider predictions given this smoothing. 
This is only one issue to be careful about, associated with the relation
between high-$k$ and small {\it difference} in $r$ effects. There are other
potential stumbling blocks associated with other limits of the coordinates. 
I think the correct way to look at this issue is to understand that sometimes 
$P(k)$ 
can be more convenient to work with than $\xi(r)$, or vice versa, and when they
appear to imply fundamentally different conclusions it is probably because you 
are trying to 
ignore the inconvenience of one of them, rather than dealing with it 
appropriately, not because they really are fundamentally different. 

Finally, since I have been hearing a lot lately in talks and conversation
the unqualified statement that
``SPT doesn't work'', which seems to blatantly contradict previous 
conclusions, it is 
useful to review the literature on this, which turns out to be very 
interesting. The short summary is that it would
be more accurate to say that ``even though SPT works very well where we need
it to, one can imagine a different Universe in which it wouldn't, so it is
fun to think about that'', or, more practically ``SPT works great up to 2 loops
but can't be pushed beyond that to give very high accuracy, so some 
improvement is desirable''. First, the old positive result one remembers 
is certainly correct, e.g., \cite{2006ApJ...651..619J} show that 1-loop SPT
gives an obvious large improvement over linear theory at all redshifts they
compute, $z=1$ and higher
(see also \cite{2012JCAP...12..013A}). 
\cite{2009PhRvD..80d3531C} confirms this result and
adds a 2-loop calculation, which again dramatically improves the results over
1-loop at $z=1$ (see also \cite{2011JCAP...08..012O}). 
Here we see, however, that while the 1-loop calculation offers
some improvement over linear theory at $z=0$, where the amplitude of the linear
power is a lot higher, the 2-loop calculation arguably actually makes things 
worse. This is intriguing, but not very relevant as there is very little volume
in the Universe near $z=0$. Things get really exciting in the 3-loop 
calculation of \cite{2014JCAP...01..010B} (if you are going to look at one
figure in another paper related to this discussion, it should be the $z=0.833$
panel of their Fig. 2), who show that, after appearing to
be converging {\it beautifully} at 1 and 2-loops at $z\sim 1$, 
the 3-loop SPT result 
is suddenly catastrophically wrong 
(this was anticipated by \cite{2014PhRvD..89b3502B}). 
\cite{2014JCAP...01..010B} explains that this
is not so much a divergence of the wavenumber integrals involved as an
explosion of the combinatoric factor in front of the terms. Apparently
this is well known behavior in quantum field theory, including even QED 
\cite{1952PhRv...85..631D}, where the series expansions are generally also 
asymptotic, not convergent. This is not considered to be a big 
fundamental problem, e.g., in identifying the issue in QED, 
\cite{1952PhRv...85..631D} was careful to state in the abstract that 
``The divergence in no way restricts the accuracy of practical calculations
that can be made with the theory'' -- the solution is simply to not calculate
beyond the point where the terms in the series are getting smaller. While this
is in practice a hundred or more terms in QED, because the coupling constant is
so small, it is clearly more of a practical problem for SPT, because, while 
the 2-loop result is great at $z\sim 1$, it 
will have larger errors than the statistical
errors of a survey like DESI \cite{2013arXiv1308.4164F}
once one pushes to high enough $k$, and clearly
simply going to 3-loops is not an option
(although, on the other hand, one
has to wonder if this will be the limitation once the unavoidable bias 
parameters are marginalized over). 
Note that \cite{2014JCAP...01..010B} show that the 2-loop calculation is pretty
good for $z$ even as low as 0.375, i.e., the really bad breakdown only 
happens at a completely irrelevantly low redshift (there is $>20$ times as
much volume in the Universe in the range $0.4<z<1.5$ as $z<0.4$, 
and of course a lot more
at higher $z$).
To be clear, I don't think this invalidates the goal of improving on SPT, 
I just think the motivation should be 
presented more accurately -- the motivation is abstract mathematical physics 
interest, i.e., applicable to Universes other than ours, and/or it is detailed
improvements on the success of SPT in our Universe. 

To conclude, the attractive fixed-point behavior dominating small scales in 
equation 
(\ref{eq:basicRG}) is a reason to be intrigued by this method, but I found that
it also contains a practical drawback: you do need to do the wavevector 
integral out to high $q$ to preserve this behavior, and therefore need to 
track the
evolution of the power spectrum out to high $q$, and straightforward 
numerical
evolution of the differential equation tends to grind to a halt with 
impossibly small steps. This problem is not insurmountable, and several other
groups have also overcome it, but it is something to be 
aware of -- you don't want to set out to evolve equation (\ref{eq:basicRG}) 
with the assumption that it will be trivial.

\acknowledgements

I thank Uro\v{s} Seljak for helpful discussions. 
 
\bibliography{cosmo,cosmo_preprints}

\end{document}